\documentclass[12pt]{article}
\usepackage{amssymb}
\DeclareMathSymbol{\ltimes}{\mathbin}{AMSb}{"6E}

\begin{document}
\hfill{\bf December 1998, IFT UWr 921/98}

\vspace{48pt}

\begin{center}{\Large \bf Deformed Quantum Relativistic Phase Spaces}
\\
{\Large \bf --
an Overview\footnote{Supported by KBN grant 2P03B13012}}
\end{center}

\vspace{18pt}
\begin{center}{\sl
 J. Lukierski\footnote{To be published in the Proceedings of
 III  International Workship ``Classical  and
 Quantum Integrable Systems", Yerevan, July 1998,
 Eds. L.D. Mardoyan, G. Pogosyan and A.N. Sissakyan, JINR Dubna
Publ. Dept.}
 \\ Institute for Theoretical Physics,\\
  University of Wroc{\l}aw,\\
pl. Maxa Borna 9, 50--205 Wroc{\l}aw, \\ Poland.
}
\end{center}
\vspace{36pt}

We describe three ways of modifying the relativistic Heisenberg
algebra - first one not linked with quantum symmetries, second
and third related with the formalism of quantum groups. The
third way is based on the identification of generalized deformed
phase space with the semidirect product of two dual Hopf
algebras describing quantum group of motions
 and the corresponding quantum Lie algebra. 
As an example the $\kappa$--deformation of relativistic
Heisenberg algebra is given, determined by $\kappa$--deformed
D=4 Poincar\'{e} symmetries.
\section{Introduction}
The ``naive" canonical quantization scheme, based on the
Heisenberg relations $(i=1,\ldots N)$
$$
\left[ \widehat{x}_{i}, \widehat{p}_{j} \right] =
i\hbar\delta_{ij}
\eqno(1.1a)
$$
$$
\left[ \widehat{x}_{i}, \widehat{x}_{j} \right] =
\left[ \widehat{p}_{i}, \widehat{p}_{j} \right] =0
\eqno(1.1b)
$$
has been applied to quantum mechanics ($N=3;$ in general $N=3n$ finite)
as well as to quantum field theory $(N=\infty)$.
The modification
of canonical quantization scheme in the presence 
 of interactions in quantum field theory is
well--known, and the difficulty with existence of canonical ET
limits causes the appearance of infinite wave function
renormalization. It appear however, that also
quantum--mechanical system in the presence of quantized
gravitational forces can not be quantized canonically. The
quantized gravitational field modifies the classical space--time
by replacing classical Riemannian geometry by 
at
 present not yet well understood quantum Riemannian geometry.
One of expected properties of such a quantized geometry is 
 the noncommutativity
of space coordinates
($[x_{i},x_{j}]\neq 0 $) [1-4]. Further
if we supplement 
the fourth Minkowski coordinate
  $x_{0} =c t$ and 
add $p_{0} = {E\over c}$, one gets the following relativistic
extension of the relations (1.1a-b)

$$
\left[ \widehat{x}_{\mu}, \widehat{p}_{\nu} \right] =
i\hbar\eta_{\mu\nu}
\eqno(1.2a)
$$
$$
\left[ \widehat{x}_{\mu}, \widehat{x}_{\nu} \right] =
\left[ \widehat{p}_{\mu}, \widehat{p}_{\nu} \right] =0
\eqno(1.2b)
$$
Again the relations (1.2a-b) will not be valid for arbitrary
small distances, because 
 taking into consideration the quantum gravity  effects
$ \left[ {x}_{\mu}, {x}_{\nu} \right] \neq
 0$ 
 [1-4]\footnote{This conditions is 
weaker than
$\left[ \widehat{x}_{i}, \widehat{x}_{j} \right] \neq0$
because one can have in relativistic case
$\left[ \widehat{x}_{i}, \widehat{x}_{j} \right] = 0$
but
$\left[ \widehat{x}_{0}, \widehat{x}_{i}^{-} \right] \neq 0$
(see also Sect. 4).}

 In this lecture I would like to review briefly different ways of
obtaining deformations of relativistic phase space 
 algebra
(1.2a-b). The distinction between different ways depends on the
role played by quantum deformations of respective symmetry
groups. We shall distinguish the following three classes of
deformations of the Heisenberg algebra (1.2a-b):
\begin{description}
\item{i)} The deformed relations remain covariant under
classical symmetry group. Such a deformation we shall call as
described in the framework of classical symmetries (see Sect. 2).
\item{ii)} The quantum phase space coordinates are the
corepresentations of quantum symmetry group and the deformed
relations are covariant under the coaction of quantum symmetries.
Such formalism is based on differential calculus on noncommutative
spaces, with the derivatives representing momenta (see Sect 3).
In particular the deformations of the relations (1.2a-b) is
obtained if we consider quantum phase space as described by the
vector corepresentation of quantum Lorentz group (deformed
Minkowski space) supplemented with noncommutative vector fields
(deformed fourmomenta).
\item{iii)} The 
 generalized quantum
relativistic phase space is described by the
semidirect product of two dual Hopf algebras, describing
respectively the quantum Poincar\'{e} group and quantum
Poincar\'{e} algebra. It appears that if 
 we 
perform the contraction by putting trivial Lorentz group generators
$(\Lambda_{\mu}^{\ \nu}=\delta_{\mu}^{\ \nu})$ one obtains the
eight-dimensional deformed relativistic phase space subalgebra,
span by the translation generators of quantum Poincar\'{e} group
and fourmomentum generators of quantum Poincar\'{e} algebra. In
particular we shall present in Sect. 4 such an example derived
from deformed $\kappa$--Poincar\'{e} symmetries. In Sect. 5 we
shall present some final remarks. 
\end{description}

\section{Deformations of Rela\-tivistic Hei\-sen\-berg 
Algebra in the
Framework of Clas\-si\-cal Re\-la\-ti\-vis\-tic Symmetries}
The first well-known proposal of noncommutativity for four
relativistic space-time coordinates is due to Snyder ([5]; see
also [6]). He
assumed that (we put $\hbar=1$):
$$
\left[ \widehat{x}_{\mu}, \widehat{x}_{\nu} \right] =
 i \ell^{2}\, M_{\mu\nu}\, ,
\eqno(2.1)$$
where $M_{\mu\nu}$ are Lorentz algebra generators, and
$$
\left[ M_{\mu\nu}, \widehat{x}_{\rho}\right]
= i\left( 
\eta_{\nu\rho} \widehat{x}_{\mu}
- 
\eta_{\mu\nu} \widehat{x}_{\rho}\right)
\eqno(2.2)
$$
In order to describe the deformation of relativistic phase space
(1.2a-b) we employ the commuting fourmomenta
($[p_{\mu},p_{\nu}]=0$). We get
$$
\begin{array}{c}
\displaystyle
M_{\mu\nu} = i
 \left(
 p_{\mu} \, 
 {\partial \over \partial p^{\nu} }
- p_{\nu} \, 
{\partial \over \partial p^{\mu} }
 \right)
\cr\cr
\displaystyle
{\widehat{x}}_{\mu} = i\left(
1 + \ell^{2}p^{2}\right)^{1\over 2}
\, {\partial \over \partial p^{\mu}} 
\, ,
\end{array}
\eqno(2.3)$$
and
$$
[ 
{\widehat{x}}_{\mu}, 
{\widehat{p}}_{\nu} 
]
= i 
\left(1 + \ell^{2} p^{2} 
\right) 
^{1 \over 2} \, \eta_{\mu\nu}\, .
\eqno(2.4)
$$
It  is clear from the relations (2.1-4) that the deformed
formalism is covariant under the \underline{classical} Lorentz
symmetries. Moreover, the relations (2.1-2) indicate that in the
momentum space one can introduce fifth momentum coordinate
$p_{5}$, satisfying the five - dimensional mass-shell condition
$p^{2}_{5} - p^{2} 
= {1 \over \ell^{2} }$
or 
$p_{5} =
 {1\over l}
 (1+ \ell^{2} p^{2} )^{1\over 2}$. It appears
that the formalism has build in classical O(3,2)
anti-de-Sitter symmetry with the deformation parameter $\ell$
equal to the inverse de-Sitter radius
in five-dimensional momentum space. This property further
was developed by Kadyshevsky 
[6] and his collaborators [7,8] as a way
 of introducing regularization
in local relativistic field theory.  

Second model which we would like to quote here has been recently
proposed by Dopplicher, Fredenhagen and Roberts (DFR) [1].
Starting from the analysis of the uncertainty relations in the
presence of quantum gravity effects (see also [2-4]) they
proposed the following model of noncommutative space-time coordinate:
$$
\left[ \widehat{x}_{\mu}, \widehat{x}_{\nu} \right]
= i\, \ell^{2}_{p}\,  \Sigma_{\mu\nu}\, ,
\eqno(2.5)
$$
where $\Sigma_{\mu\nu} = - \Sigma_{\nu\mu}$ is the two-tensor,
commuting with the coordinates $\widehat{x}_{\mu}$. It appears
that the relation (2.5) can be supplemented by the 
\underline{classical} relations
$[p_{\mu},p_{\nu}]=0$,
$[\widehat{x}_{\mu},p_{u}]=i\eta_{\mu\nu}$. The DFR phase space
relations are covariant under the classical Lorentz
symmetries, provided that the Lorentz generators $M_{u\nu}$ contain
additional term rotating the tensor components $\Sigma_{\mu\nu}$
$$
\left[
M_{\mu\nu}, \Sigma_{\rho\tau}\right]
= 
i\left(
\eta_{\nu\rho} \, \Sigma_{\mu\tau} -
\eta_{\mu\rho} \, \Sigma_{\nu\tau}
+
\eta_{\mu\tau} \, \Sigma_{\nu\rho}
-
\eta_{\nu\tau} \, \Sigma_{\mu\rho}\right)\, .
\eqno(2.6)
$$
Moreover, because $\Sigma_{\mu\nu}$ are
$x_{\mu}$-independent translation-invariant operator, we see that
the DFR formalism is invariant under the classical D=4
Poincar\'{e} transformations.

The relation (2.5) can be written for any D, but only in 
D=2 one can use standard Poincar\'{e} generators, because the
central two-tensor $\Sigma_{\mu\nu}$ will be described by a scalar:
$$
\Sigma^{\rm D=2}_{\mu\nu} = \eta_{\mu\nu} \cdot \Sigma
\, .
\eqno(2.7)
$$
Interestingly, if we consider D=2 Euclidean space, and
corresponding  (2+1)--dimensional Galilei group, the relation
(2.5) describes the second central extension of classical
(2+1)--dimensional algebra [9,10], with the nonrelativistic
coordinates described by boost generators.

The DFR deformations is very mild, and its dynamical
consequences can be calculated explicitly (for D=4 see [1,11];
for Euclidean D=2 case see [12]).

It should be pointed out that both deformations presented in
this section introduce besides the phase space generators $Y_{A}
=
(\widehat{x}_{\mu},\widehat{p}_{\mu})$ also additional
generators $U_{r}$ (Lorentz generators $M_{\mu\nu}$ or tensorial
central charges $\Sigma_{\mu\nu}$), which form together consistent
associative algebra:
$$\begin{array}{c}
\displaystyle
[ Y_{A},Y_{B} ] = \Omega_{AB} (Y,U)\, ,
\cr\cr
\displaystyle
[Y_{A}, U_{r} ] = \Omega_{Ar}(Y,U)\, ,
\cr\cr
[U_{r},U_{s}] = \Omega_{rs}(Y,U) \, .
\end{array}\eqno(2.8)$$

The generalized quantum symplectic structure 
($ \Omega_{AB}=- \Omega_{BA},  \Omega_{Ar},  \Omega_{rs}=-\Omega_{sr}$) satisfy
Jacobi identities. In Snyder case the algebra (2.8) is the
 anti-de-Sitter algebra in five-momentum space 
constrained
 by the mass-shell condition $p^{2}_{5}-p^{2}={1\over \ell^{2}}$. 
 
 In general case discussing the geometric and algebraic nature
of deformed phase space algebra (2.8) one should ask three questions:
\begin{description}
\item{i)} Under which symmetries the relations (2.8) are covariant?
 In the following two paragraphs we shall consider the examples of
deformations covariant under quantum symmetries, described by
noncommutative and noncocommutative Hopf algebra playing the
role of
symplectomorphism group.
\item{ii)} which reality conditions for the generators
$Y_{A},U_{r}$ can be consistently imposed?
\item{iii)} which are the Hilbert space (or C$^{\star}$-algebra)
realizations of the deformed quantum phase space algebra (2.8)?
\end{description}

In this brief review we shall discuss the deformations in purely
algebraic framework, and we shall be mainly concerned with the
answer to the first question.
\section{Deformations of Heisenberg Algebra Co\-va\-riant under the
Quantum Group Transformations}

For canonical case, describing relativistic quantum mechanics 
(see (1.2a-b)  the basic
 set of relations (2.8) is reduced to the first one with $\Omega_{AB}=
 \pmatrix{0 &\eta_{\mu\nu}\cr
 -\eta_{\mu\nu} & 0 } $,
 and the symplectomorphism group is described by Sp(8;R). The
symplectomorphism algebra sp(8;R) contains the diagonal
subalgebra O(3,1) describing the classical Lorentz
transformations in space - time and four-momentum sectors.

First 
step 
in considering deformed Heisenberg algebra (1.2) is to consider
deformed Lorentz symmetry.
At present there are well-known all possible quantum Lorentz
symmetries, in the form of real $\star$-Hopf algebra [13,14]. The
problem which has not been yet solved is the 
embeddings of all these quantum Lorentz groups into the  possible
quantum symplectomorphism groups described by quantum deformations
of Sp(8;R). 

At this point we would like to observe, that Heisenberg algebra (1.1)
 can be also presented as an algebra of creation and
annihilation operators $a_{i}={1\over \sqrt{2}} (\widehat{x}_{i}
+ i\widehat{p}_{i})$,
 $a_{i}^{+}={1\over \sqrt{2}} (\widehat{x}_{i}
- i\widehat{p}_{i})$ 
satisfying well-known relation (we put $\hbar=1$ )
$$ \begin{array}{c}
\displaystyle
[a^{+}_{i},a_{j}] =\delta_{ij}
\cr\cr
\displaystyle
[a_{i}, a_{j}]=[a^{+}_{i}, a^{+}_{j}]=0
\end{array}
\eqno(3.1)
$$
If $i=1,\ldots, N$ the symplectomorphism groups is Sp(2N), but
there is a subgroups $U(N)\subset Sp(2N)$ of holomorphic
transformations 
$a'_{i}=U_{i}^{\ j}a_{j}$ 
where $U_{i}^{j} \in U(N)$.
The problem of deformation of relations (3.1) consistent with
well-known Drinfeld-Jumbo deformation $U_{q}(N)$ has been solved
by Woronowicz and Pusz ([15]; see also [16]). The set of
q-deformed commutation relations looks as follows
$([A,B]_{q}\equiv {A}B -qBA)$:
$$
\begin{array}{ll}
\displaystyle
[a_{i},a^{+}_{j}]_{q}=0  &i>j
\cr\cr
\displaystyle
[a'_{i},a^{+}_{j}]_{q^{2}}=1
+
\left(q^{2}-1\right) \sum\limits^{i-1}_{k=0} a^{+}_{k}a_{k}
& i=j
\cr\cr
[a_{i},a_{j}]_{q}=
[a^{+}_{i},a^{+}_{j}]_{q^{-1}}=0
&i<j
\end{array}\eqno(3.2)
$$
It appears that the 
n-dimensional modules 
($a_{1},\ldots , a_{N}$) and 
($a^{+}_{1},\ldots , a^{+}_{N}$) form the corepresentation of
quantum holomorphic symplectomorphism group $U_{q}(N)$\footnote{
For the definition of $U_{q}(N)$ see [17,18].}.
This construction has been further generalized to the
supersymmetric set of bosonic and fermionic creation and
annihilation operators [19] and to the case of 
quantum covariance 
group described by a quantum R-matrix [16,20].

In order to describe the quantum deformation of the relativistic
phase space (1.2a-b) one should perform the following steps:

\begin{description}
\item{i)} Specify the quantum deformation O(3,1)$^{(q)}$ of
Lorentz group O(3,1), which
 only in particular cases 
  can be described in terms of
quantum R-matrices for the spinorial Lorentz group SL(2;c).
\item{ii)} Determine the algebra of quantum Minkowski space 
${\cal{M}}^{(q)}_{4}$, with four generators of algebra describing
the fundamental representation of O$^{(q)}$(3,1).
\item{(iii)} Define the quantum fourmomenta $p_{\mu}$ by
 noncommutative 
derivatives (noncommutative vector fields) 
acting on the functions on 
${\cal{M}}^{(q)}_{4}$.
\end{description}

It appears that depending on the type of deformations, we need
to consider besides the generators 
$Y_{A}=(\widehat{x}_{\mu},\widehat{p}_{\mu})$
also additional generators 
 $U_{r}$ (see (2.8)). We shall mention here the following two cases:

 \begin{description}
 \item{i)}
 Drinfeld-Jimbo deformation of Lorentz group.
 
 Such a case was elaborated in detail by Munich group [21-23].
The relations (2.8) are realized if we supplement 7 additional
generators $U_{r} =(M_{\mu\nu},\Lambda)$, where
$M_{\mu\nu}$ are the Lorentz generators, and $\Lambda$ describe
the scaling generator. An interesting property of this scheme
is the classical nature of time coordinate (i.e. ${\widehat{x}}_{0}$ can be
chosen an central element of 
${\cal{M}}^{(q)}_{4}$).
\item{ii)} Podle\'{s} class of deformations of Poincar\'{e}
 group, providing 8-dimensional quantum phase space.
 
 It is known [24,25] that the Drinfeld-Jimbo
deformation of Lorentz group can not be extended to 
quantum 
Poincar\'{e} group. Only the deformations of Lorentz group with
quantum R-matrix satisfying \hbox{$R^{2}=1$} permits the extension to quantum
Poincar\'{e} group 
${\cal{P}}^{(q)}_{4}$
 [26].
The translation sector of 
${\cal{P}}^{(q)}_{4}$ describes the generators 
$\widehat{x}_{\mu}$ of 
${\cal{M}}^{(q)}_{4}$:
\end{description}
$$
(R- 1)^{\ \ \mu\nu}_{\rho\tau}
\left( {\widehat{x}}^{\rho}{\widehat{x}}^{\tau} - Z^{\rho\tau}_{\ \ \nu}
{\widehat{x}}^{\nu} + T^{\rho\tau} \right) = 0
\eqno(3.3)
$$
where $R$ is quantum O(3,1) matrix satisfying $R^{2}=1$ and 
$Z_{\mu\nu}^{\ \ \rho}$, 
$C_{n\nu}$ are numerical dimensionless coefficients,
satisfying suitably set of constraints (see [26]). The
fourdimensional differential calculus, with four independent basic
 one-forms and corresponding four dual derivatives, are obtained
if the parameters in eq. (3.3) satisfy additional
 conditions [27]. In such a case the space-time noncommutativity
(3.4) can be supplemented by the following remaining relations
of deformed algebra (1.2a-b)

$$
p_{u}{\widehat{x}}^{\nu} - R^{\nu\rho}_{\ \ \mu\tau}
{\widehat{x}}^{\tau} p_{\rho} =
\delta_{\mu}^{\ \nu} 
+ Z^{\nu\rho}_{\ \ \mu} p_{\rho}
\eqno(3.4a)
$$

$$
p_{u}p_{\nu} - R^{\tau\rho}_{\ \ \nu\mu} p_{\rho}p_{\tau}
= 0
\eqno(3.4b)
$$
where $p_{\nu}$ can be identified with the noncommutative derivatives.

\section{Deformed Phase Space as Heisenberg Double of Quantum
Space--Time Symmetries}
In previous paragraphs we described deformed relativistic phase
space as a deformed Heisenberg algebra covariant under classical
(Sect. 2) or quantum (Sect. 3) relativistic symmetries. In such an
approach to ``deformed physics" the basic object is a quantum
phase algebra, and the quantum symmetries in form of respective
Hopf algebras describe its covariance properties. It should be
stressed that the assumptions of covariance of deformed phase
space algebra (deformed Heisenberg algebra) under classical or
quantum symmetries restricts substantially all possible class of
deformations. 

In this Section we shall consider the scheme in which primary
object is the quantum symmetry. The deformed relativistic phase
space will be obtained directly from a dual pair of Hopf
algebras, describing quantum Poincar\'{e} group
${\cal{P}}^{(q)}_{4}$
and quantum Poincar\'{e} algebra
${{P}}^{(q)}_{4}$. The realizations describing the
noncommutativity of coordinates and momenta will be obtained by
introducing the semidirect product 
${\cal{P}}^{(q)}_{4} \ltimes
{{P}}^{(q)}_{4}$, or so--called Heisenberg double [28,29].

If we have two dual Hopf algebras $H$, $\widetilde{H}$, with
duality generated by inner
 nondegenerate product
 $<\widetilde{a}, a>
  (a \in H; \widetilde{a} \in \widetilde{H})$, the Heisenberg
double ${\cal{H}}(H)$ 
is the algebra $H \otimes \widetilde{H}$ with the cross
multiplication given by the relations:
$$
\widetilde{a} \circ a = a_{(1)} <{\widetilde{a}}_{(1)},
a_{(2)}>
{\widetilde{a}}_{(2)}
\eqno(4.1)
$$
where we use the 
 notation $\Delta(a) = a_{(1)} \otimes a_{(2)}$ etc.
In particular one can apply this definition to the
multiplication of two Abelian Hopf algebras describing
translation sector of classical Poincar\'{e} group
$$
\left( \widetilde{H} = \{ x^{\mu}\};
[x^{\mu}, x^{\nu} ] = 0,
\qquad \Delta(x^{k}) =x^{k}\otimes 1
+ 1 \otimes x^{k}\right)
$$
and fourmomentum sector of classical Poincar\'{e} algebra
$(H=\{ p_{k};[p^{k},p^{v}]=0$,
 $\Delta(p_{k})=p_{k}\otimes 1 + 1\otimes p_{k})$.
 Introducing the Planck constant $\hbar$ into the duality relation
 $$
 <x^{\mu}, p_{\ \nu}> = i\hbar \delta^{\mu}_{\ \nu}
 \eqno(4.2a)
 $$
 one obtains from (4.1)
 $$
 x^{\mu}\cdot p_{\nu} 
 = p_{\nu} <1,1>x^{\mu} + 1 \cdot <x^{\mu},p_{\nu}> \cdot 1
 \eqno(4.2b)
 $$
 Because $<1,1>=1$, substituting (4.2a) we obtain from 
 (4.2a) the classical relativistic Heisenberg algebra (1.2a-b).
Such a scheme can be generalized to the case of any quantum -
deformed relativistic symmetries. In this lecture I shell
provide the example of so--called $\kappa$--deformations of D=4
relativistic symmetries. 

We choose the following realization of the $\kappa$-Poincar\'{e}
algebra in bicrossproduct basis [2,8]
 $(\mu,\nu,\lambda,\rho=0,1,2,3;i,j=1,2,3$ and $g^{\mu\nu} =
g_{\mu\nu} = (-1,1,1,1))$

-algebra sector
$$
\begin{array}{l}
[P_{\mu},P_{\nu}]=0
\cr\cr
[M_{\mu\nu},M_{\lambda\sigma}] =
i\hbar (g_{\mu\sigma} M_{\nu\lambda}
+ g_{\nu\lambda}M_{\mu\sigma}
-g_{\mu\lambda}M_{\nu\sigma}
-g_{\mu\sigma}M_{\mu\lambda}
\cr\cr
[M_{ij},P_{\mu}]=-i\hbar
(g_{i\mu} P_{j}-g_{j\mu}P_{i})
\cr\cr
[M_{i0},P_{0}]= i\hbar P_{i}
\cr\cr
[M_{i0},P_{j}] = i\delta_{ij}
\left( \hbar^{2}
 \kappa \sinh 
\left( {P_{0}\over \hbar
\kappa}\right)
 e^{-{P_{0}\over \hbar\kappa}}
+ {1\over 2\kappa} \vec{P}^{2} \right)
- {i\over \kappa}
P_{i}P_{j}
\end{array}\eqno(4.3)
$$

-coalgebra sector
$$\begin{array}{l}
\Delta(M_{ij}) = M_{ij}\otimes I +
I \otimes M_{ij}
\cr\cr
\cr\cr
\Delta(M_{k0}) = M_{k0}\otimes
 e^{-{P_{0}\over \hbar\kappa}}
+
I \otimes M_{k0}
+{1\over \hbar \kappa} M_{kl} \otimes P_{l}
\cr\cr
\Delta(P_{0}) =
P_{0}\otimes 
 I   +  I \otimes P_{0}
\cr\cr
\Delta(P_{k}) =
P_{k}\otimes 
 e^{-{P_{0}\over \hbar\kappa}} + I\otimes P_{k}
 \end{array}
 \eqno(4.4)
 $$
where $\hbar$ denotes Planck`s constant and $\kappa$ deformation
parameter. The formulas for the antipode and 
 counit are omitted because they are not essential for this construction.
 
 Using the following duality relations, extending (4.2)
 $$
 <x^{\mu},P_{\nu}> = i\hbar \delta^{\mu}_{\nu}
 \quad
 <\Lambda^{\mu}_{\ \nu}, M_{\alpha \beta}>
 = i \hbar \left( \delta^{\mu}_{\alpha} g_{\nu\beta} 
 - \delta^{\mu}_{\beta}g_{\nu\alpha}\right)
 \eqno(4.5)
 $$
 we obtain the commutation relations defining
$\kappa$--Poincar\'{e} group [3,8] in the form 

- algebra sector

$$
\begin{array}{l}
[x^{\mu},x^{\nu}] = {i\over \kappa}
\left( \delta^{\mu}_{0} x^{\nu} -
 \delta^{\nu}_{0} x^{\mu}\right)
 \cr\cr
 [\Lambda^{\mu}_{\nu},x^{\lambda}]
 = -{i\over \kappa}
 \left( 
 \left( \Lambda^{\mu}_{0} - \delta^{\mu}_{0}
  \right)\Lambda^{\lambda}_{\nu} +
 \left(
 \Lambda^{0}_{\nu} - \delta^{0}_{\nu}\right)
 g^{\mu\lambda}\right)
 \cr\cr
 [\Lambda^{\mu}_{\nu}, \Lambda^{0}_{\mu}] = 0
 \end{array}
 \eqno(4.6)
 $$
 
 -coalgebra sector
 $$
 \begin{array}{l}
 \Delta(x^{\mu}) =
 \Lambda^{\mu}_{\ \alpha} \otimes x^{\alpha} 
 + x^{\mu} \otimes I
 \cr\cr
 \Delta(\Lambda^{\mu}_{\ \nu}) =
 \Lambda^{\mu}_{\ \alpha} \otimes
 \Lambda^{\alpha}_{\ \nu}
 \end{array}
 \eqno(4.7)
 $$
 The commutation relations (4.3) and (4.6) one can supplemented
by the following relations obtained from (4.1), (4.4) and (4.7)

- cross relations:

$$
\begin{array}{l}
[P_{k},x_{l}] = i \hbar \delta_{kl}
\qquad \qquad
[P_{0},x_{0}] = i \hbar 
\cr\cr
[P_{k},x_{0}] = -{i\over \kappa} P_{k}
\qquad
\qquad
[P_{0},x_{l}] = 0
\cr\cr
[P_{\mu},\Lambda^{\alpha}_{\beta}]=0
\cr\cr
[M_{\alpha \beta},\Lambda^{\mu}_{\ \nu} ]
= i\hbar 
\left(
\delta^{\mu}_{\beta} \Lambda_{\alpha\nu} 
- \delta^{\mu}_{\alpha} \Lambda_{\beta\nu} 
\right)
\cr\cr
[M_{\alpha \beta},x^{\mu} ]
= i\hbar 
\left(
\delta^{\mu}_{\beta} x_{\alpha} 
- \delta^{\mu}_{\alpha} x_{\beta}\right) 
+
 {i\over \kappa}
\left( \delta^{0}_{\beta} M_{\alpha}^{\ \mu}
- \delta^{0}_{\alpha}M_{\beta}^{\ \mu} 
 \right)
\end{array}
\eqno(4.8)
$$
where
$$
M_{\alpha}^{\ \mu} = g^{\mu\rho} M_{\alpha\rho}
\qquad
M_{\ \alpha}^{ \mu} = g^{\mu\rho} M_{\rho\alpha}
$$

The relations (4.3), (4.6) and (4.8) give us the Heisenberg double
${\cal{H}}({\cal{P}}_{k})$ of $\kappa$-Poincar\'{e} 
 algebra firstly presented in [32].
In order to define the $\kappa$--deformed phase space we should
consider subalgebra of ${\cal{H}}({\cal{P}}_{k})$ given by the
following commutation relations:
$$
\begin{array}{ll}
[x_{k},x_{l}] = 0 \qquad\qquad
 &
 [P_{\mu},P_{\nu}] = 0
\cr\cr
[x_{0},x_{k}] = {i \over \kappa} x_{k}
\cr\cr
[x_{k},P_{l}] = i\hbar \delta_{kl} \qquad\qquad
&
[x_{k},P_{0}] = 0
\cr\cr
[x_{0},P_{l}] ={ i \over \kappa}P_{l}
\qquad\qquad
&
[x_{0},P_{0}] =- i\hbar 
\end{array}
\eqno(4.9)
$$
For $\kappa \to \infty$ we get the standard nondeformed phase space
satisfying the Heisenberg commutation relations (1.2a-b).

It should be stressed that the $\kappa$--deformed phase space
(4.9) is not a Hopf algebra, however, separately in the
translation sector (generators $x_{\mu}$) and fourmomentum
sector (generators $P_{\mu}$) one can read-off from the
relations (4.4) and (4.7) (in eq. (4.7) after contraction
$\Lambda^{\beta}_{\beta} \to \delta^{4}_{\alpha}$) the coalgebraic
structure. We obtain that in $\kappa$--deformed phase space one
can add two coordinates and two momenta as follows:
$$
x^{(1+2)}_{\mu} = x^{(1)}_{\mu} + x^{(2)}_{\mu}
\eqno(4.10a)
$$
$$
P^{(1+2)}_{0} = P^{(1)}_{0} + P^{(2)}_{0}
\eqno(4.10b)
$$
$$
\vec{P}^{(1+2)} = \vec{P}^{(1)}
e^{- {P^{(2)}_{0}\over \hbar\kappa}} + \vec{P}^{(2)}
\eqno(4.11c)
$$
We see that due to the $\kappa$--deformation there is
  modified the addition law of two three--momenta.
 
 Analogous Heisenberg double construction can be performed for
any quantum Poincar\'{e} group listed by Podle\'{s} and
Woronowicz [26].
It should be mentioned however, that before
solving such a task
 we should obtain the
complete list of quantum Poincar\'{e} algebras, dual to quantum
groups from [26] - the list which at present is yet not known.
\section{Final Remarks}
The considerations of deformed quantum phase spaces goes along the
following three lines:
\begin{itemize}
\item{i)} Mathematical consequences of deforming space--time 
symmetries. In such a way one obtains the set of possible algebraic
schemes, which we outlined in this talk.
\item{ii)} Investigation of dynamical consequences of modifying
Einstein gravity scheme (e.g. superstring theory) on Heisenberg
uncertainty relations (see e.g. [32,33])
\item{iii)} Considering different ``gedenken-experiments" for the
measurement of distances and time intervals in quantum theory,
which include into analysis the quantum property of the observer
(reference system) (see e.g. [34-36]). It appears that the ``classical"
 measuring devices, with infinite mass as well as definite
position and velocity are the idealizations in the description
of measuring process
  which is not
realistic. It appears that  introducing ``realistic"
measurement theory in quantum mechanics we are bound to modify
the standard Heisenberg uncertainty relations, with Planck length
$l_{p} = {\hbar\over m_{p}c} \simeq 10^{-33}$ cm playing a
crucial role in the modifications.
\end{itemize}

We see that the future modified models of quantum mechanics
should have on one side  solid 
mathematical and dynamical bases, on the other should be
consistent with ``realistic" measurement theory. A modest
effort in such a direction has been presented in [36], where
there are given arguments based on ``realistic" measurement
theory for the $\kappa$--deformation of relativistic phase
space, presented in Sect. 4. 
At present, however, the choice of
the deformation which is the most physical one is an open question. 

\end{document}